\documentclass[]{raa}
\usepackage{deluxetable}
\usepackage{amssymb}            
\usepackage{graphicx,times}
\usepackage{natbib}
\usepackage{color}
\usepackage[figuresleft]{rotating}

\providecommand{\bm}[1]{\mbox{\bol
dmath$#1$\unboldmath}}

\begin{document}

\title{Capability of Searching for Kilonova Associated with a Short Gamma-ray Burst by SVOM}

\volnopage{}
	\setcounter{page}{1}

	\author{J. Wang\inst{1,2,*}, L. P. Xin\inst{1}, Y. L. Qiu\inst{1}, L. Lan\inst{1}, 
	W. J. Xie\inst{1}, Z. P. Jin\inst{3}, J. Y. Wei\inst{1}
	}
 \institute{Key Laboratory of Space Astronomy and Technology, National Astronomical Observatories, Chinese Academy of Sciences, Beijing 100101, China. {wj@nao.cas.cn}. \and Guangxi Key Laboratory for Relativistic Astrophysics, School of Physical Science and Technology, Guangxi University, Nanning 530004, China, .
 \and Purple Mount Astronomical Observatories, Chinese Academy of Sciences, Nanjing 100101, China. 
}

\abstract{ 
In spite of the importance of
studying the cosmic generation of heavy elements through the r-process,
the detection of kilonova resulted from a merger of neutron star binaries is still a challenge task.
In this paper, we show that the Visible Telescope (VT) onboard the on-going SVOM space mission
is powerful for identifying kilonova candidates associated with short gamma-ray bursts (SGRBs) 
up to a distance of 600Mpc.
A significant color variation, turn blue and then turn red, is revealed 
by calculating the light curves in both red and blue channels of VT by a linear combination of an afterglow and an associated kilonova. The maximum color variation
is as high as $\sim0.5-1$ mag, which is far larger than the small 
photometry error of $\sim0.2$ mag of VT for a point source with a brightness of 23 mag. Up to a distance of 600Mpc, $\sim1-2$ kilonova candidates per year are predicted to 
be identified by VT. 
\keywords{gamma-ray burst: general --- stars: neutron --- instrumentation: photometers --- telescopes}
}

 \authorrunning{Wang et al.}            
 \titlerunning{Searching for kilonova by SVOM}  
 \maketitle

\section{Introduction}

The merger of neutron star binaries in the Hubble timescale is predicted to manifest
itself as 
gravitational wave (GW) radiation (e.g., Abbott et al. 2020), short-duration 
gamma-ray burst (SGRB, $T_{90}<2$s, e.g., Kouveliotou et a. 1993), and associated kilonova 
(e.g., Li \& Paczynski 1998;
Eichler et al. 1989; Freiburghaus et al. 1999; Rosswog et al. 1999; Perego et al. 2014;
Just et al. 2015). The kilonova is powered by the 
radioative decay of isotopes of the heavy elements that is assembled by the so-called 
rapid neutron capture (r-process) nucleosynthesis in the matter expelled by the merger
(e.g., Burbrige et al. 1957; Barnes \& Kasen 2013; Barnes etal. 2016; Kasen et al. 2013, 2017; Metzger et al. 2010; Metzger 2019; Chen et al. 2024; Korobkin et al. 2012).

Although SGRBs have been frequently detected by past and on-going GRB missions (see review in 
Berger 2014), 
the detection 
of the associated kilonova is still a hard task at the current stage. In fact, only two have been confirmed by spectroscopy in literature. The first case, AT~2017gfo associated with 
a weak SGRB GRB~170817A, was found by the Swope Supernova 
Survey during 
a campaign of searching for the electromagnetic counterpart of GW~170817 
discovered by the LIGO-Virgo experiments (e.g., Abbott et al. 2017a, b; Andreoni et al. 2017; Arcavi et al. 2017; Kilpatrick et al. 2017; Covino et al. 2017; Cowperthwaiteet al. 2017; Drout et al. 2017; Evans et al. 2017; Smartt et al. 2017; Tanaka et al. 2017; Goldstein et al. 2017). 
The identification of the kilonova was confirmed by comparing its spectroscopic sequence obtained by large telescopes with the spectral models (e.g., Pian et al. 2017; 
Shappee et al. 2017). The second one is GRB~230307A, a long-duration GRB at $z=0.065$, 
in which Levan et al. (2024)
recently identified an emission-line of tellurium (atomic mass $A=130$) and very red spectral-energy distribution from the mid-infrared
spectroscopy and imaging taken by James Webb Space Telescope at dozens of days after the trigger. 

In addition to the two spectorscopically confirmed cases, by
modeling the multi wavelength light curves,
a batch of kilonova candidates have been 
identified due to an excess of near-infrared (NIR) emission at a couple of days after the 
GRBs' trigger 
(e.g., Berger et al. 2013; Fan et al. 2013; Tanvir et al. 2013; Jin et al. 2015, 2016, 2020, 2021; 
Yang et al. 2015; Troja et al. 2018, 2019; Lamb et al. 2019; Rastinejad et al. 2022; Zhu et al. 2023). 

In this paper, we demonstrate that the Visible Telescope (VT) onboard the SVOM satellite 
enables us to easily identify kilonova candidates up to a distance of $\sim600$Mpc according to 
their strong NIR excess. 
The main reason for this fact is that the red channel of VT has quite deep sensitivity up to 
a wavelength of 1$\mu$m since the lack of the strong NIR atmospheric emission in space.  
The paper is organized as follows. Section 2 briefly describes the 
payloads of the SVOM mission, especially the capability of VT. 
The calculations of the optical light curves observed by VT 
are presented in 
Section 3. Section 4 gives the results and implication.

\section{Instruments onboard SVOM}

SVOM, launched in 2024, June 22, is a Chinese-French space mission dedicated to the detection and study of GRBs. 
We refer the readers to Atteia et al. (2022) and the white paper given by Wei et al. 
(2016) for the details.

There are four onboard instruments. The wide-field soft $\gamma$-ray imager ECLAIRs 
and Gamma-Ray Monitor (GRM) are designed to observe GRB prompt emission in
4–150 keV and 15–5000 keV energy bands, respectively.
With a field-of-view (FoV) of 2sr and a sensitivity of 
$7.2\times10^{-10}\ \mathrm{erg\ s^{-1}\ cm^{-1}}$ (5$\sigma$ detection level in an exposure of 1000 s), 
a total of 60-70 GRBs per year can be triggered by ECLAIRs
(Godet et al. 2014; Cordier et al. 2015). In addition, with a detection area of 
200$\mathrm{cm^{2}}$ of each GRD module, $\sim90$ GRBs are expected to be detected per year by GRM.

The narrow-field Micro-channel X-ray Telescope (MXT, Gotz et al. 2014)
and Visible Telescope (VT) are responsible for follow-up
observations of the afterglows in X-ray and optical wavelength, respectively.
VT is a Ritchey-Chretien telescope with a 44 cm
diameter and a $f$-ratio of 9. It has a FOV of about 26$\times$26 $\mathrm{arcmin^2}$, covering the ECLAIR’s error box in most cases. 
The limiting magnitude is down to $m_V = 22.5$mag for a 300 s exposure. 
With a dichroic beam splitter, VT works in two channels, one in
blue and the other in red, simultaneously. The blue channel has a wavelength range 
from 0.4 to 0.65 $\mu$m, and the red one from 0.65 to 1.0 $\mu$m. 
Each channel is equipped with a 2k$\times$4k E2V frame transfer CCD, in which
a back-illuminated thick CCD is used for the red channel to enhance the quantum efficiency (QE).

The left panel in Figure 1 shows the total throughput curves of the two channels, along
with the corresponding transmittance of the filters and the
QE of the two CCDs. More detailed description on the calibration and determination of 
the throughput curves can be found in Qiu et al. (in preparation). With the throughput curves, the 
limiting magnitudes in the two channels are calculated by the dedicated
Exposure Calculator of VT\footnote{The calculator can be 
visited at the website of SVOM Science User Support System through the link of https://svom-gwacn.cn/gp/tools/CalcExptimeVT.action.} according to the simplified ``CCD'' equation (Mortara \&
Fowler 1981; Gullixson 1992; Merline \& Howell 1995 and see NOAO/KPNO CCD instrument manuals)
\begin{equation}
  \mathrm{\frac{S}{N}}\simeq\frac{\eta N_\star t}{\sqrt{\eta N_\star t+n_{\mathrm{pix}}(\eta N_{\mathrm{b}} t + N_{\mathrm{d}}t+\mathrm{RN}^2})}
\end{equation}
where $N_\star$, $N_\mathrm{b}$ and $N_\mathrm{d}$ are the photon rate from the source, the photon rate per pixel 
from sky background, and dark current per pixel, respectively.
$t$ is the exposure time and RN the readout noise of each
pixel. $n_{\mathrm{pix}}$ and $\eta$ are the number of pixels occupied by a single
star and the system total throughput, respectively.
The calculated limiting magnitudes (see Section 3.3 for the definition of magnitude of SVOM/VT) 
in both channels are plotted in the right panel of Figure 1 
as a function of exposure time for a powerlaw 
$f_\nu\propto\nu^{-1.3}$ (typical of a GRB, see below) at two significance levels of $3\sigma$ and $5\sigma$. Basically, for an exposure of 300 seconds, the limiting magnitudes are 23.4 and 23.1
for the blue and red channels, respectively, at a significance level of 3$\sigma$. The 
corresponding values degrade to 22.9 and 22.6 at a significance level of 5$\sigma$. 

\rm

\begin{figure}
   \centering
   \begin{minipage}{0.49\linewidth}
      \centering
      \includegraphics[width=1.0\linewidth]{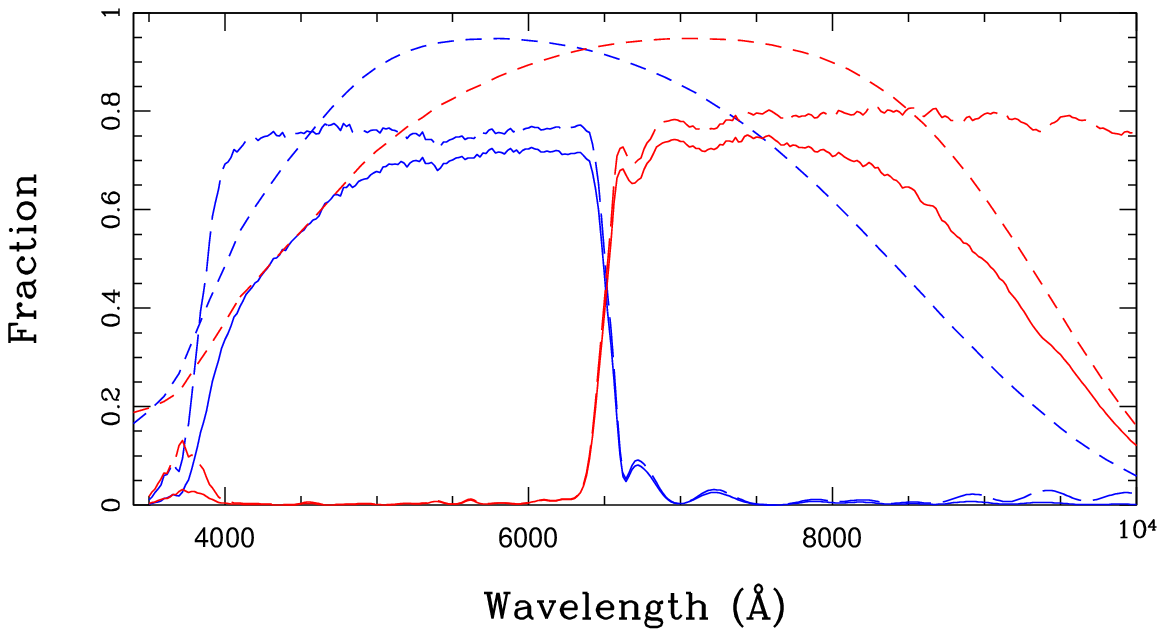}
   \end{minipage}
   \begin{minipage}{0.49\linewidth}
      \centering
      \includegraphics[width=1.0\linewidth]{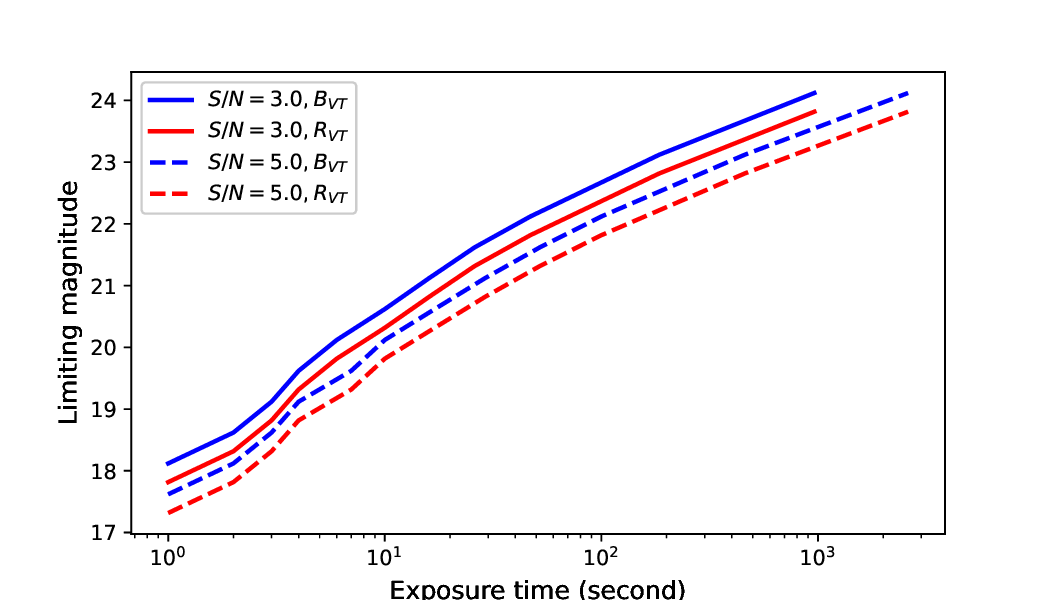}
   \end{minipage}
   \caption{
   \it Left panel: \rm The total throughput curves (the solid line), 
   transmittance of the filters (the long-dashed line)
   and QE of the CCDs (the short-dashed line) of VT as a
   function of wavelength. The blue and red channels are 
   denoted by blue and red colors, respectively. \it Right panel: 
   The predicted limiting magnitudes in both VT channels 
   plotted against exposure time at two significance 
   levels, i.e., $3\sigma$ and $5\sigma$.
}
\label{Fig1}
\end{figure}

\section{Calculation of Theoretical Light Curves}

To examine the capability of identifying kilonova candidates,
we calculate a set of 
light curves recorded in both VT channels by a linear combination of a SGRB's afterglow
and an associated kilonova. 

\subsection{Afterglow from SGRB}

The evolution of the luminosity of an afterglow is calculated by adopting the simple and widely used 
powerlaw: $L^{\mathrm{grb}}_\nu\propto\nu^{-\alpha}t^{-\beta}$. 
The values of $\alpha$ and $\beta$ are fixed to be 
1.3 and 1.2, respectively, for a late afterglow (i.e., at about 1 day after a trigger and 
ignore the possible jet break effect) throughout the paper. 
The coefficient of the powerlaw is determined from
a sample of the normalized light curves of SGRBs (e.g., Nicuesa Guelbenzu et al. 2012). Specifically, 
the brightness in the $R_c-$band at 1 day is adopted to be from 29 to 25 mag 
in the $z=1$ reference frame.

\subsection{Emission from Kilonova}

The spectral model grid developed by Kasen et al. (2017)
\footnote{https://github.com/dnkasen/Kasen\_Kilonova\_Models\_2017} is adopted by us to calculate 
the specific luminosity $L^{\mathrm{kn}}_\nu$ of a given kilonova at a given time.
In the model, the mergers of two neutron stars are simulated in general relativity,
in which the material is ejected by two distinct ways. One with a high velocity of 0.2-0.3c is
resulted from the dynamical expelling of the 
matter from the surface of the approaching stars due to the tidal force, along with the  
squeezing of the matter at the interface. The another low-velocity (0.05-0.1c) ejection is caused 
by the wind from the accretion disk formed around the central remnant after the merger. 
The ejecta is assumed to be spherical symmetric with a density profile of $\propto\upsilon^{-1}$ and 
$\propto\upsilon^{-10}$ in the inner and outer layers, respectively. The emission resulted from 
the decay of radioative r-process isotopes is  
calculated by solving the time-dependent radiative transfer equation the under local thermodynamic 
equilibrium, in which millions of bound-bound transitions are considered to obtain the 
opacities, including all lanthanides.

The model grid provides spectra from ultraviolet 
to infrared at every 0.1 days after a merger. In addition to the fixed 
exponents of the inner and outer density profile, the parameters of the model grid are:
the ejecta mass $0.001\leq m/M_\odot\leq0.1$, the kinetic velocity of the ejecta 
$0.03\leq v_k/c\leq0.3$ and the mass
fraction of lanthanides $10^{-9}\leq X_{\mathrm{lan}}\leq10^{-1}$.

\subsection{Predicted Light Curves Recorded by VT}

With the time resolved spectra of both afterglow and kilonova described above,
the light curves recorded in the two VT channels (denoted by 
$B_{\mathrm{VT}}$ and $R_\mathrm{VT}$ for the blue and red channels, respectively) 
are calculated according to the definition of the AB magnitude (Fukugita et al. 1996): \rm
\begin{equation}
   m_{\mathrm{AB}}= -2.5\log\frac{\int f_\nu S_\nu \mathrm{d}\ln\nu}{\int S_\nu \mathrm{d}\ln\nu}-48.6
\end{equation}
where $S_\nu$ is the total throughput at frequency $\nu$ as given in Figure 1.
$f_\nu$ is the specific flux density of an object in the unit of 
$\mathrm{erg\ s^{-1}\ cm^{-2}\ Hz^{-1}}$, and is calculated from the predicted luminosity $L_\nu$ as 
$f_\nu=L_\nu/4\pi d^2$, where $d$ is the distance.

\section{Results}

Figure 2 compares the predicted light curves between an afterglow and a kilonova both 
at a distance 
of 40Mpc (the upper panel) and 600Mpc (the lower panel). In each panel, the boundaries of the afterglow are inferred for the two extreme cases, i.e., a strong afterglow with 
$m_{R_c}=25$ mag and a weak afterglow with $m_{R_c}=29$ mag, where $m_{R_c}$ is the brightness 
in the $R_c-$band at 1 day in the $z=1$ reference frame. The kilonova light curves are obtained from 
the model spectra with $m=0.05M_\odot$, $v_k=0.3c$ and $X_{\mathrm{lan}}=10^{-4}$.

Three facts can be learned from the figure.
At first, the VT two channels reproduce the well known infrared excess of a kilonova at late epoch of
$\sim1$day,
which will be addressed below in more details. 
Secondly, in the red channel, the light curve at $\sim1$day after a trigger can be dominated by a kilonova if the
associated afterglow is quite weak (or off-axis). 
Finally, with the limiting magnitude of $\sim$22.5mag, 
a kilonova candidate can be identified in the VT red channel at a distance as far as $\sim$600Mpc. 

Figure 3 shows the predicted temporal evolution of $B_\mathrm{VT}-R_\mathrm{VT}$ color obtained by VT 
for the whole kilonova spectral model grid given in Kasen et al. (2017). \rm  The fiducal model is again adopted to be 
the one with $m=0.05M_\odot$, $v_k=0.3c$ and $X_{\mathrm{lan}}=10^{-4}$.
In addition to the strong and weak afterglow cases, 
the cases with an intermediate afterglow level 
with $m_{R_c}=27$ mag are shown in the middle column. One can see from the figure 
a significant color variation in almost all the cases, in which the light curves become blue at early 
epoch, and transform to red at late epoch. The weaker the associated afterglow, the larger the value of 
color difference $\Delta(B_\mathrm{VT}-R_\mathrm{VT})$ will be. The maximum $\Delta(B_\mathrm{VT}-R_\mathrm{VT})$ is as high 
as $\sim 0.5-1$mag. It is noted that $\Delta(B_\mathrm{VT}-R_\mathrm{VT})$ is in fact independent on the spectral 
shape of the associated afterglow as long as its spectral index maintains a constant.   
This revealed trend is consistent with the theoretical evolution of 
a kilonova, where the kilonova spectrum is dominated by the 'blue' (light r-process) component at 
the beginning, and then by the 'red' (heavy r-process) component after a couple of days.  

We argue that there is an agreement between the spectral model and the 
dependence of the calculated color evolution on kilonova parameters. At first, the top row in Figure 3
shows that a brighter kilonova that results in a redder spectrum is produced by a larger 
eject mass, which is consistent with the scaling law for the characteristic luminosity 
$L\propto m^{0.35}v^{0.65}\kappa^{-0.65}$, i.e., Eq. (3) in Kasen et al. (2017), where 
$\kappa$ is the opacity being sensitive to $X_{\mathrm{lan}}$: larger the $X_{\mathrm{lan}}$,
lager the $\kappa$ will be.
Secondly, the Eq. (2) in Kasen et al. (2017) leads to a duration of kilonova 
$t\propto m^{1/2}v^{-1/2}\kappa^{1/2}$. This scaling law implies a shorter-lasting kilonova for 
a higher velocity, which can be learned from the middle row in the figure. Finally, 
the bottom row in the figure shows the dependence on $X_\mathrm{lan}$. \rm
For the cases 
with a large $X_\mathrm{lan}>10^{-2}$, the corresponding large opacity causes a kilonova 
emission primarily emerged in infrared (i.e., the heavy r-process component), 
which results in a reduced effect on the optical color.
On the contrary, a large variation of optical color can be revealed in the 
cases with small $X_\mathrm{lan}<10^{-4}$, in which the kilonova emission in 
the optical bands is dominated by the light r-process component that decays and cools 
with time.

\begin{figure}
   \centering
   \includegraphics[width=10cm]{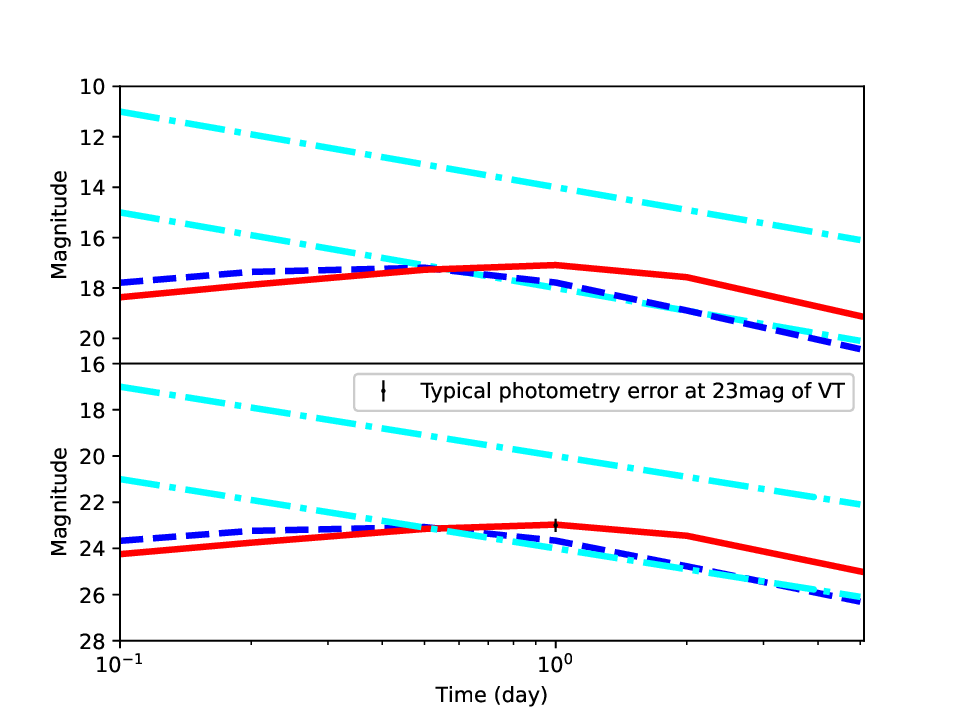}
   \caption{
   Calculated light curves of an afterglow and a kilonova for a distance of 40Mpc (upper panel) 
   and 600Mpc (lower panel). In each panel, the two cyan dot-dashed lines correspond to the light
   curves of an 
   afterglow with $R_c=25$ and 29mag, the brightness in the $R_c-$band at 1 day in the $z=1$ reference frame. The red-solid and blue-dashed lines are the kilonova 
   light curves recorded in VT red and blue channels, respectively. See the text for the 
   details of calculations. The vertical black line in the lower panel
   marks the typical magnitude error of $\sim$0.2~mag measured by VT for a source with a brightness of 23mag. 
}
\label{Fig1}
\end{figure}

\begin{figure}
   \centering
   \includegraphics[width=\columnwidth]{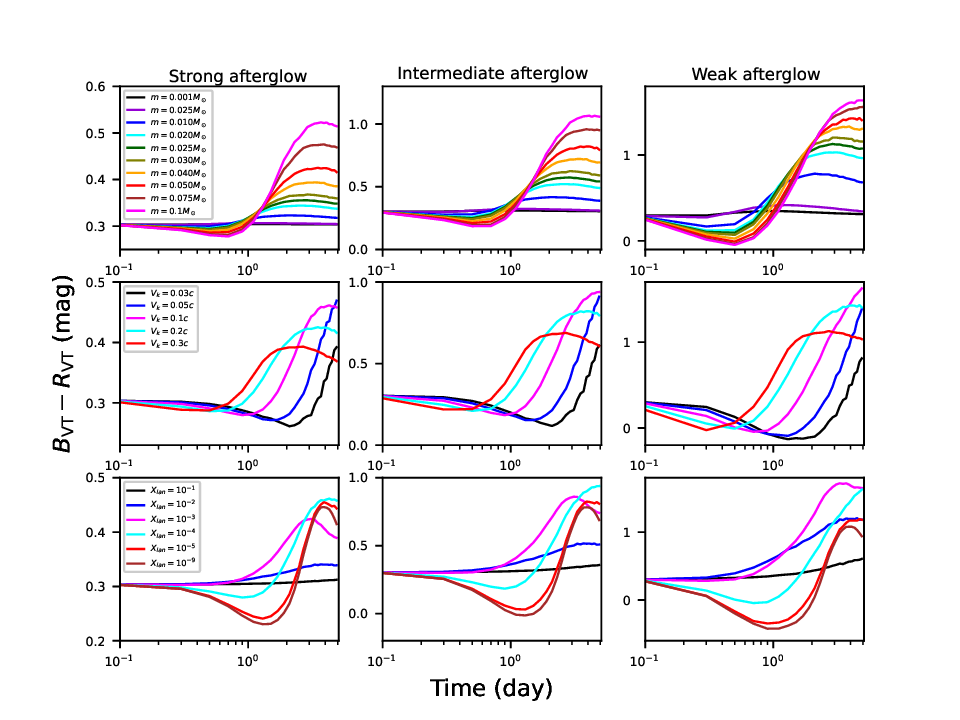}
   \caption{
   Evolution of the VT $B_\mathrm{VT}-R_\mathrm{VT}$ color assessed by a linear combination of a SGRB's afterglow
   and a kilonova. The three columns correspond to different afterglow levels. The dependence on
   the ejecta mass ($m$), kinetic velocity of the ejecta ($v_k$) and mass
   fraction of lanthanides $X_{\mathrm{lan}}$ are displayed by the top, middle and bottom rows, respectively. See the main text for the details of the calculations. 
}
\label{Fig1}
\end{figure}

Based on our calculations,
we conclude that VT is powerful for identifying a kilonova associated with a SGRB 
through the measured color variation of the afterglow. In fact,
VT has typical photometry errors of $\sim0.01$ and $\sim0.2$ mag for a point source with brightness of 16 and 23 mag, respectively, which enables us to identify a kilonova candidate at a distance up to 600Mpc by VT. At this distance, in the case with a strong associated 
afterglow, the $B_\mathrm{VT}-R_\mathrm{VT}$ color is predicted to change by $\sim$0.4 mag
for the fiducal kilonova model with $m=0.05M_\odot$, $v_k=0.3c$ and $X_{\mathrm{lan}}=10^{-4}$
(i.e., the red lines in Figure 3). This predicted color change is slightly larger than or 
comparable to the error of the measured color of $\sqrt{2}\times0.2\approx0.3$ mag.

%

\section{Conclusion and Implications}

The light curves in both red and blue channels of VT onboard the SVOM satellite
are predicted by a linear combination of an afterglow of a SGRB and an associated kilonova.
The predicted light curves show a significant color variation with the maximum value 
as high as $\sim0.5-1$ mag. With the detection limit and accuracy, VT is therefore  
powerful for identifying kilonova candidates up to a distance of 600Mpc

We estimate the detection rate of kilonova by VT as follows. 
Based on the recent BATSE GRB catalog\footnote{https://gammaray.nsstc.nasa.gov/batse/grb/catalog/current/.}, there are 500 SGRBs with $T_{90}<2$s among the 2041 events, which yields a SGRB 
fraction of $\sim1/4$ in the BATSE sample. Since the GRM onboard SVOM has 
comparable energy range with the BASTE, $\sim20$ SGRBs in total are predicted to be detected per year 
by GRM. Among the 20 SGRBs, there are $\sim1-2$ SGRBs with a luminosity distance up to 600Mpc
by assuming the SGRBs detected by GRM follow the redshift distribution of SGRBs given in Berger (2014),
which leads to a kilonova detection rate of $\sim1-2\mathrm{yr^{-1}}$ by VT onboard the SVOM.

\begin{acknowledgements}
The authors would like to thank the anonymous referees for his/her careful review and 
helpful comments.
This study is supported by the Strategic
Pioneer Program on Space Science, Chinese Academy of Sciences,
grants XDB0550401. JW is supported by the National Natural Science Foundation of
China (Grants No. 12173009),
by the Natural Science Foundation of Guangxi
(2020GXNSFDA238018) and by the Bagui Young Scholars Program. LL is supported by 
the National Postdoctoral Program for Innovative Talents (grant No. GZB20230765).

\end{acknowledgements}

\label{lastpage}

\end{document}